\documentclass[12pt]{article}
\usepackage[super,compress]{cite}
\usepackage{amsmath}
\usepackage{amssymb}
\usepackage{graphicx}
\usepackage{float}
\begin{document}

\begin{center}
{\bf High energy nucleus-nucleus collision and halo radii
in different approaches
of Glauber theory.
}

Yu.M. Shabelski and A.G. Shuvaev \\

\vspace{.5cm}

Petersburg Nuclear Physics Institute, Kurchatov National
Research Center\\
Gatchina, St. Petersburg 188300, Russia\\
\vskip 0.9 truecm
E-mail: shabelsk@thd.pnpi.spb.ru\\
E-mail: shuvaev@thd.pnpi.spb.ru

\end{center}

\vspace{1.2cm}

\begin{abstract}
\noindent
The complete Glauber calculation
of the differential cross sections of $^{12}$C--$^{12}$C
and halo nuclei on $^{12}$C scattering was performed
using the previously proposed
in Refs.~\cite{Shabelski:2021iqk, Shabelski:2022xkw}
method of generating function.
The results are different as compared with the similar calculations
in the optical model
and the rigid target approximation. The halo nuclei radii
extracted from the scattering data via the complete Glauber
analysis come out to be larger than those obtained
in the approximate approaches

\end{abstract}

\section{Introduction}

In the previous papers \cite{Shabelski:2021iqk, Shabelski:2022xkw}
we suggest a novel approache to the Glauber theory
allowing one to analytically account
for all the Glauber diagrams
for nucleus\textendash{}nucleus scattering without additional
approximations.
It relies on the employment of the generation function
that produces the complete Glauber amplitudes.
In the present paper we compare it with more simple methods,
in particular, in the Section 2 the differential cross section
of $^{12}$C -- $^{12}$C scattering evaluated in our approach is
compared with those obtained in the optical model and in the rigid
target model.
We also study the effect due to incorporation
of the real part of the strong scattering amplitude
apart from the imaginary one
and its interference with Coulomb interaction.

We carry out the calculation
of the reaction cross sections for the halo nuclei
$^{8}$B, $^{11}$Li, $^{11}$Be, $^{14}$Be and
using the experimental data on their scattering
on $^{12}$C we have extracted the halo radii.
The results are presented in Section 3.
One has to note that the agreement with the experimental cross sections
is achieved in complete Glauber for a bit larger nuclear radii than
in the optical model.

\section{Elastic $^{12}$C\textendash{}$^{12}$C scattering
in different approaches}

The first calculations of the elastic nucleus\textendash{}nucleus
scattering applying the generation function method has been published
in our previous paper \cite{Shabelski:2022xkw}.
Here we briefly recall the main points of the formalism.
The generating function $Z(u,v)$ provides the amplitude
of the elastic scattering of the incident nucleus
$A$ on the fixed target nucleus $B$,
\begin{eqnarray}
F_{AB}^{el}(q)\,&=&\,\frac{ik}{2\pi}\int d^{\,2}b\,e^{iqb}
\bigl[1\,-\,S_{AB}(b)\bigr], \nonumber \\
\label{der}
S_{AB}(b)\,&=&\,\frac 1{Z(0,0)}
\frac{\partial^A}{\partial u^A}
\frac{\partial^B}{\partial v^B}\,
Z(u,v)\biggl|_{u=v=0},
\end{eqnarray}
where $q$ is the transferred momentum and $k$ is the mean
nucleon momentum in nucleus $A$,
the two-dimensional impact
vector $b$ lies in the transverse plain to the momentum~$k$.

The elastic amplitude is simply related to the total cross section
through the optical theorem,
$$
\sigma_{AB}^{tot}\,=\,\frac{4\pi}{k}
\mathrm{Im}F_{AB}^{el}(q=0)\,=\,
2\int\!\! d^2b\,\bigl[1\,-\,S_{AB}(b)\bigr].
$$
The difference between the total cross section
and the integrated elastic cross section,
$$
\sigma_{AB}^{el}\,=\,
\int\!\! d^2b\,\bigl[1\,-\,S_{AB}(b)\bigr]^2,
$$
yields the reaction cross section,
$$
\sigma_{AB}^{r}\,=\,\sigma_{AB}^{tot}\,-\,\sigma_{AB}^{el}
\,=\,
\int\!\! d^2b\,\bigl[1\,-\,S_{AB}^2(b)\bigr].
$$
Although it is so-called interaction
cross section rather than the reaction one
that is experimentally measured, the difference
between them is estimated to be no more than
2-3\%
\cite{Novikov:2013zdw}.

The closed expression for the function $Z(u,v)$ has been obtained
in Ref.~\cite{Shabelski:2021iqk}
\begin{eqnarray}
\label{Zuv}
Z(u,v)\,&=&\,e^{W_y(u,v)},~~~~~~~~
z_y\,=\,1-\frac 12 \frac{\sigma_{NN}^{tot}}{a^2},
\\
\label{Wy}
W_y(u,v)\,&=&\, \frac 1{a^2}\int d^{\,2}x\,
\ln\bigl(\!\!
\sum\limits_{M\le A,N\le B}
\frac{z_y^{M\, N}}{M!N!}
\bigl[a^2 u\rho_A^\bot(x-b)\bigr]^M
\bigl[a^2 v\rho_B^\bot(x)\bigr]^N
\bigr).
\end{eqnarray}
The transverse densities entering this formula are expressed
through the three dimensional nucleon distributions in
the colliding nuclei,
$$
\rho_{A,B}^\bot(x_\perp)\,=\,
\int d z\,\rho_{A,B}(z,x_\perp),~~~
\int d^2 x_\perp\,\rho_{A,B}^\bot(x_\perp)\,=\,1,
$$
$\sigma_{NN}^{tot}$ is the total nucleon-nucleon
cross section, the value $a^2 = 2\pi\beta$
is related to the slope of the elastic nucleon-nucleon
amplitude, see Eq.(\ref{fNN}) below.

The function $W_y(u,v)$ (\ref{Wy})
goes as the series built of the overlaps,
\begin{equation}
\label{tmn}
t_{m,n}(b)=
\frac 1{a^2}\int d^{\,2}x\,
\bigl[a^2\rho_A^\bot(x-b)\bigr]^m\,
\bigl[a^2\rho_B^\bot(x)\bigr]^n,
\end{equation}
with $m\le A$ and $n\le B$.
Keeping only the lowest $m=n=1$ term
we arrive at the well-known optical approximation~\cite{Czyz:1969jg}
\begin{equation}
\label{t11}
F(A,B)\,=\,
-\frac 12 \,\sigma_{NN}^{tot}\,
T_{AB}(b),~~~~
T_{AB}(b)\,=\,A\, B\, t_{1,1}(b).
\end{equation}
Another known approximation
is the rigid target (or projectile)
approximation~\cite{Bialas:1977pd, Alkhazov:1977ur}.
It requires one density, say, $\rho_A^\bot(x)$,
to be kept in the formula (\ref{Zuv}) only in the linear
order, permitting at the same time any powers
of $\rho_B^\bot(x)$.
It yields the generating function
$$
Z(u,v)\,=\,e^{v + u T_{rg}(v,b)},~~~~
T_{rg}(v,b)\,=\,\sum_{n=0}^\infty \frac 1{n!}\,
t_{1,n}(b)\,v^n,
$$
producing for $B\gg 1$
\begin{equation}
\label{rt}
S_{AB}(b)\,=\,\bigl[T_{rg}(b)\bigr]^A,~~~~
T_{rg}(b)\,=\,\int d^{\,2}x\,\rho_A^\bot(x-b)\,
e^{-\frac 12 \sigma_{NN}^{tot}\rho_B^\bot(x)}.
\end{equation}

The complete Glauber amplitude implies all the pieces
(\ref{tmn}) to be included in the generating function
before the derivatives (\ref{der}) are taken.
For relatively light nuclei, $A,B\lesssim 15$,
it can be done straightforwardly.

In this paper the nucleon density
has been taken in a simple
Gaussian parameterizations well suited for light nuclei,
\begin{equation}
\label{Gauss}
\rho(r)\,=\,\rho_0\,e^{-\frac{r^2}{a_c^2}}.
\end{equation}
This form differs from that used in our previous
papers~\cite{Shabelski:2021iqk, Shabelski:2022xkw}.
The change is motivated by exotic halo nuclei
analysis in the next section since it is this form
that is employed for the density of the core.
The value $a_c$ is expressed through the mean square
nuclear radius, $a_c=\sqrt{3/2}R_{rms}$.
The nucleon-nucleon elastic scattering amplitude
is taken the same as before
except for the ratio of the real to imaginary parts,
$\varepsilon$, added,
\begin{equation}
\label{fNN}
f_{NN}^{el}(q)\,=\,\frac{ik}{4\pi}\sigma_{NN}^{tot}
(1-i\varepsilon)\,
e^{-\frac 12\beta q^2}.
\end{equation}
For the energy around 1000~MeV per projectile
nucleon the total nucleon-nucleon cross section
and the slope value (averaged over $pp$ and $pn$
interaction) are~\cite{Alkhazovi:2011ty,Horiuchi:2006ga}
\begin{equation}
\label{param}
\sigma_{NN}^{tot}=43~\mathrm{mb},~~~
\beta = 0.2~\mathrm{fm}^2.
\end{equation}

With these parameters and the overlap functions (\ref{tmn})
evaluated for the distribution (\ref{Gauss})
one gets the generating function and the amplitude (\ref{der}).
The mean square radius $R_{rms}$ has been adjusted to match
the experimental interaction (reaction) cross section
$\sigma_{^{12}\mathrm{C}-^{12}\mathrm{C}}^r=853\pm 6$~mb
at the energy about 1~GeV
per nucleon~\cite{Ozawa:2000gx, Ozawa:2001hb}.
It turns out to be
sufficiently dependent on the approximation,
the optical model gives $R_{rms}=2.19$~fm, in the rigid target
approximation $R_{rms}=2.27$~fm, whereas the complete Glauber
calculation results into $R_{rms}=2.44$~fm.
This is a consequence of the fact that for a given radius
the optical model cross section would be the largest,
the additional screen  corrections make it lower.

Taking the last radius value, $R_{rms}=2.44$~fm, we have calculated
the differential cross sections
of the elastic $^{12}$C\textendash{}$^{12}$C scattering
in the three above approaches for $\varepsilon=0$.
The curves in Fig.1~(left panel) demonstrate
a significant difference especially in the diffractive minima
positions. The left curve is for the optical model,
the next one is for the rigid target approximation
and the rightmost curve stands for the complete Glauber
calculation.

The situation shown in Fig.1~(right panel) is opposite
in the sense that the radius value is separately adjusted
for each curve to have the common cross section,
$\sigma_{^{12}\mathrm{C}-^{12}\mathrm{C}}^r=853$~mb,
for all of them.
Now the curves
are very close at the interval between
$q^2=0$ and the first minimum, then
the difference increases with the transferred
momentum growth.
\begin{figure}[H]
\hskip -1.5cm
\includegraphics[width=.6\hsize]{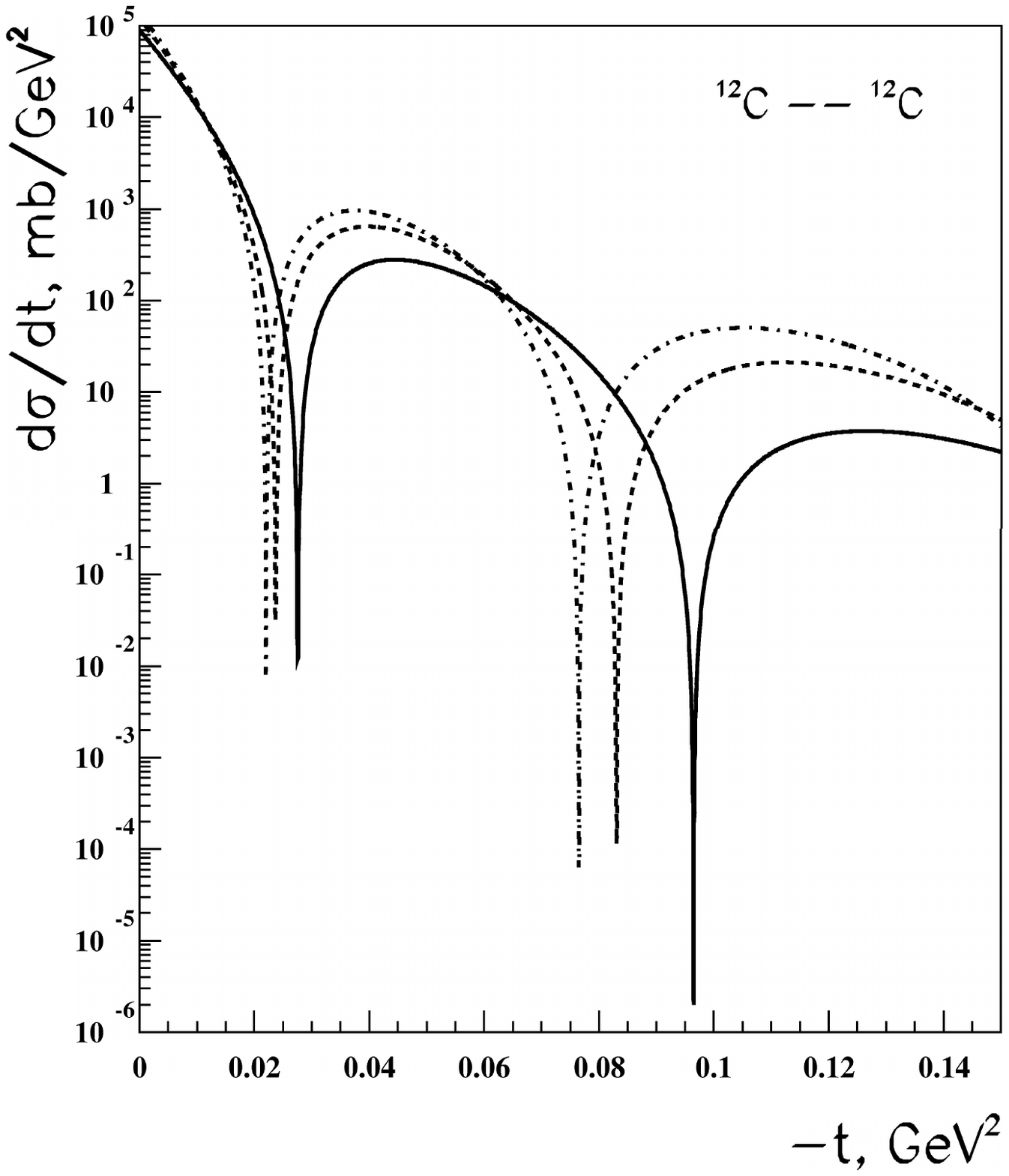}
\includegraphics[width=.6\hsize]{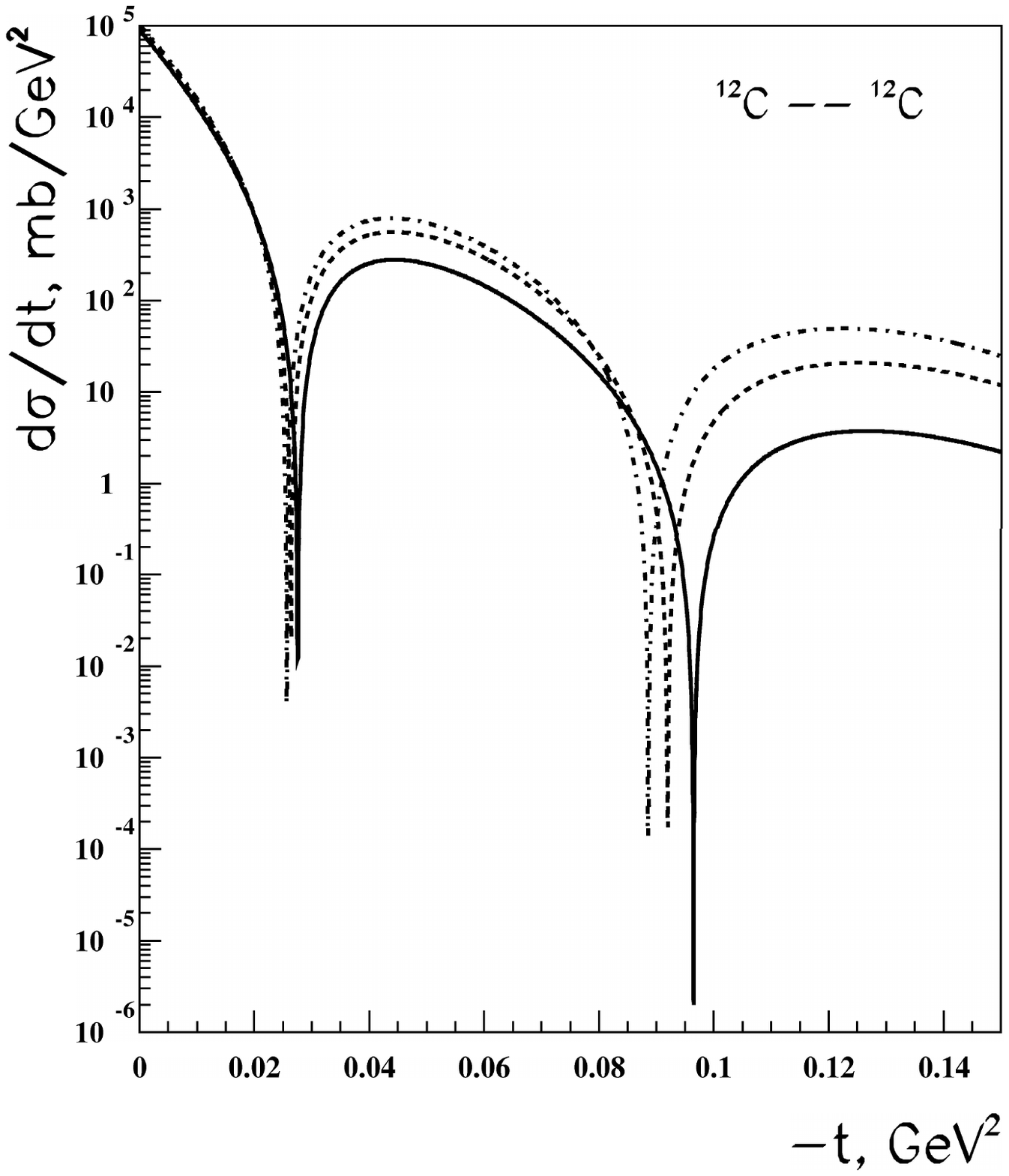}
\vskip -2.cm
\caption{\footnotesize
Left panel: The differential cross sections
of the elastic $^{12}$C\textendash{}$^{12}$C scattering
obtained in the optical model (dash-dotted line),
rigid target approximations (dotted line)
and in the complete Glauber (solid line)
for $R_{rms}= 2.44$~fm and $\varepsilon=0$
for all three curves.\\
Right panel:
The differential cross sections
of the elastic $^{12}$C\textendash{}$^{12}$C scattering
obtained in the optical model with $R_{rms}= 2.19$~fm
(dash-dotted line),
rigid target approximations with $R_{rms}= 2.27$~fm
(dotted line) dashed
and in the complete Glauber
for $R_{rms}= 2.44$~fm (solid line).
The cross section
$\sigma_{^{12}\mathrm{C}-^{12}\mathrm{C}}^r=853$~mb,
is equal for all lines,
$\varepsilon=0$.}
\end{figure}

The effect coming from the real part of the strong
interaction amplitude (\ref{fNN}) is presented in Fig.2
for the value~\cite{Horiuchi:2006ga} $\varepsilon=-0.275$,
along with its interference with Coulomb interaction.
The differential cross section in
one photon exchange approximation reads
$$
\frac{d\sigma}{dt}\,=\,\frac{\pi}{k^2}
\bigl|f_C(q)\,+\,F_{AB}^{el}(q)\bigr|^2,
$$
where the Coulomb amplitude is
$$
f_C(q)\,=\,-M_{C}Z_C^2 e^2\,\frac{\rho(q)^2}{q^2},
~~~\rho(r)\,=\,\int d^3r\,e^{iqr}\rho(r),
$$
$M_C$ and $Z_C$ being the mass and the charge number
of $^{12}$C nucleus. The Coulomb amplitude is real
and directly interplays with the real part of $F_{AB}^{el}$.

There are three curves in Fig.2 -- the differential
$^{12}$C\textendash{}$^{12}$C cross section evaluated
for $\varepsilon=0$ and without Coulomb corrections
(the same as the solid curves in Fig.1),
then the curve with the Coulomb corrections, but without
the real part of the scattering amplitude and for
the real part and Coulomb contribution combined.

As can be seen from the Fig.2
the curves are discernible only in the diffractive minima neighborhoods.
The pure Coulomb
can be separated out from its mixture with the real part only near
the first minimum, where
its contribution is several times
smaller than that due to the real part,
whereas their difference at the second minimum
is of three order of magnitude.
This is the reason why
the first order Coulomb contribution
seems to be sufficient in this treatment
as the next corrections would be irrelevant.

\begin{figure}[H]
\vskip -4.cm
\includegraphics[width=.6\hsize]{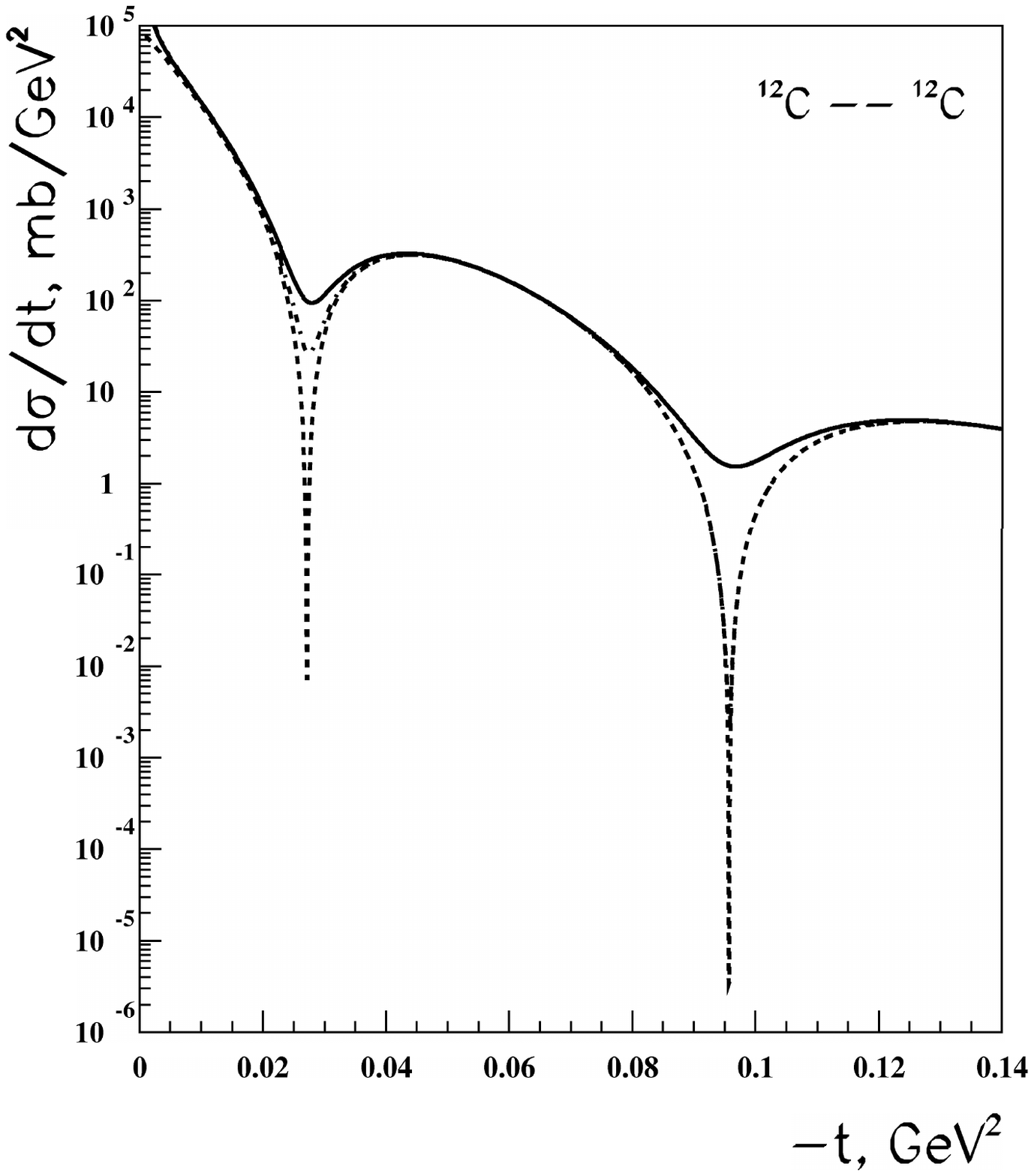}
\vskip -2.cm
\caption{\footnotesize
The differential cross sections
of the elastic $^{12}$C\textendash{}$^{12}$C scattering
in the complete Glauber approach for $\varepsilon=0$
without Coulomb contribution (dashed line)
with the Coulomb contribution (dash-dotted line)
and with $\varepsilon=-0.275$ plus Coulomb
contribution (solid line).}
\end{figure}

\section{Halo nuclei}

The standard treatment of halo nuclei assumes their
density to be the sum
\begin{equation}
\label{halo}
\rho(r)\,=\,N_c\rho_c(r)\,+\,N_v\rho_v(r)
\end{equation}
of the core including $N_c$ nucleons and surrounding
it halo with $N_v$ valence nucleons.
Throughout our analysis, the core density was parameterized
as a Gaussian function~(\ref{Gauss}), where Gaussian width,
$a_c$, is related to
the root-mean-square core radius, $R_c$, as $a_c = \sqrt{2/3} R_c$.
Depending on the shell structure, there are three commonly
used parameterizations of the halo density~\cite{Hassan:2015dfa}
\begin{equation}
\label{params}
\begin{array}{ccc}
  \rho_v^G(r)\,=\,\frac 1{\pi^{\frac 32}a_G^3}e^{-\frac{r^2}{a_G^2}},~~~&
  ~\rho_v^O(r)\,=\,\frac 2{3\pi^{\frac 32}a_O^5}\,r^2e^{-\frac{r^2}{a_O^2}},~~~&
  ~\rho_v^{2S}(r)\,=\,\frac 2{3\pi^{\frac 32}a_{2S}^3}
   \left(\frac{r^2}{a_{2S}^2}-\frac 32\right)^2
  e^{-\frac{r^2}{a_{2S}^2}},
 \\
 a_G=\sqrt{2/3}\,R_v,  & a_O=\sqrt{2/5}\,R_v, & a_{2S}=\sqrt{2/7}\,R_v,
\end{array}
\end{equation}
the mean square radius
of the halo, $R_v$, being a single parameter.

Parameter of the core density function, $a_c$, was determined
by the comparison of experimental data on the cross section
of the elastic scattering of the corresponding to the core
of exotic nucleus on $^{12}$C, Ref.~\cite{Ozawa:2000gx,Ozawa:2001hb},
with the reaction (interaction) cross section calculated
using the method outlined in previous section.
With the known core radii
and the densities (\ref{halo}),(\ref{params})
(normalized to unity)
we make the complete Glauber calculation
of the elastic cross sections
for the scattering of the exotic nuclei on $^{12}$C.
It allows to extract the halo radii $R_v$ from
the experimental data~\cite{Ozawa:2000gx,Ozawa:2001hb}
for each particular halo parametrization.
The results are shown in the Table 1.

Table 1. Mean square radii of the halo nuclei
extracted from the cross sections of their scattering
on $^{12}$C target. The core radii are chosen to match
the cross sections of the would-be core nuclei
scattering on $^{12}$C. The experimental data are
taken from
\cite{Ozawa:2000gx, Ozawa:2001hb}.
\begin{table}[H]
\centering
\begin{tabular}{|l|c|c||c|c|c|c|}
\hline
~~Nuclear structure&\multicolumn{2}{c||}{~Interaction}&
\multicolumn{4}{c|}{~Mean square radius, fm~}\\
\cline{4-7}
~~(core + halo)&\multicolumn{2}{c||}{cross section, mb}
& core &\multicolumn{3}{c|}{types of halo}\\
\cline{2-3}\cline{5-7}
 & core & halo&  &~~G~~ & ~~O~~ &2S \\
\hline
\hline
~~$^{8}$B $\to$ $^{7}$Be + p &738 $\pm$ 9 &784 $\pm$ 14 &2.47& 2.95 & 2.95 & 2.93 \\
 & & 798 $\pm$ 6 & & 3.36 & 3.36 & 3.31\\
\hline
~$^{11}$Be $\to$ $^{10}$Be + n &813 $\pm$ 10 &942 $\pm$ 8 & 2.44& 6.21 & 5.45 & 5.46 \\
\hline
~$^{11}$Li $\to$ $^{9}$Li + 2n&796 $\pm$ 6 &1040 $\pm$ 60 & 2.47& 5.51 & 5.35 & 5.36 \\
\hline
~$^{14}$Be $\to$ $^{12}$Be + 2n &927 $\pm$ 18 &1139 $\pm$ 90 &2.70& 5.42 & 5.53 & 5.55 \\
\hline
\end{tabular}
\end{table}
\noindent
Both the core and the halo radii are about 10\% larger
if compared with the similar calculations made
in the optical model~\cite{Hassan:2015dfa}.
It is interesting to note that the necessity to increase
the density radius was indicated in
Refs.~\cite{Al-Khalili:1996ugd,Al-Khalili:1996our},
though because of the different origin --
the more complicated halo structure.

\section{Conclusion}

The results of the Glauber calculations are found to be rather sensitive
to the approximation used. A sizable difference between various approaches
show up in the differential elastic scattering cross section especially
when it is evaluated with the common value of the nuclear density radius.
The curves become more close if the
reaction cross section is fixed instead of the radius,
which value is then separately adjusted within a given approximation.
In this case the radius obtained
with the complete Glauber calculation turns out to be larger
than that in the optical model. This also holds for the exotic nuclei.
The complete Glauber yields both the core and the halo radii
exceeding the optical model results.
Although the difference is not so much it is systematic.
The reason is in the screening corrections which reduce
the cross section and, therefore, require the radius
to be increased.

We are grateful to M.G. Ryskin and I.S. Novikov for useful discussions.


\begin{thebibliography}{**}

\bibitem{Shabelski:2021iqk}
Y.~M.~Shabelski and A.~G.~Shuvaev,
{\em ``Generating function for nucleus-nucleus
scattering amplitudes in Glauber theory},
Phys. Rev. C \textbf{104}, no.6, 064607 (2021)
[arXiv:2104.04943 [hep-ph]].

\bibitem{Shabelski:2022xkw}
Y.~M.~Shabelski and A.~G.~Shuvaev,
{\em Differential elastic nucleus\textendash{}nucleus
scattering in complete Glauber theory},
Mod. Phys. Lett. A \textbf{37}, no.13, 2250081 (2022)
[arXiv:2201.06270 [nucl-th]].

\bibitem{Czyz:1969jg}
W.~Czyz and L.~C.~Maximon,
{\em High-energy, small angle elastic scattering
of strongly interacting composite particles},
Annals Phys. \textbf{52}, 59 (1969).

\bibitem{Bialas:1977pd}
A.~Bialas, M.~Bleszynski and W.~Czyz,
{\em Relation Between the Glauber Model
and Classical Probability Calculus},
Acta Phys. Pol. B \textbf{8}, 389 (1977).

\bibitem{Novikov:2013zdw}
I.~S.~Novikov and Y.~Shabelski,
{\em Complete Glauber calculations of reaction
and interaction cross sections for light-ion collisions},
Phys. Atom. Nucl. \textbf{78}, no.8, 951-955 (2015)
[arXiv:1302.3930 [nucl-th]].

\bibitem{Alkhazov:1977ur}
G.~D.~Alkhazov, T.~Bauer, R.~Bertini, L.~Bimbot,
O.~Bing, A.~Boudard, G.~Bruge,
H.~Catz, A.~Chaumeaux, P.~Couvert, \textit{et al.}
{\em Elastic and Inelastic Scattering of 1.37-GeV alpha Particles
from Ca-40, Ca-42, Ca-44, Ca-48},
Nucl. Phys. A \textbf{280}, 365 (1977).

\bibitem{Alkhazovi:2011ty}
G.~D.~Alkhazovi, Y.~Shabelski and I.~S.~Novikov,
{\em Nuclear Radii of Unstable Nuclei},
Int. J. Mod. Phys. E \textbf{20}, 583-627 (2011)
[arXiv:1101.4717 [nucl-th]].

\bibitem{Horiuchi:2006ga}
W.~Horiuchi, Y.~Suzuki, B.~Abu-Ibrahim and A.~Kohama,
{\em Systematic analysis of reaction cross-sections
of carbon isotopes},
Phys. Rev. C \textbf{75}, 044607 (2007)
[erratum: Phys. Rev. C \textbf{76}, 039903 (2007)]
[arXiv:nucl-th/0612029 [nucl-th]].

\bibitem{Alkhazov:1977rg}
G.~D.~Alkhazov,
{\em Elastic Scattering of High-Energy Protons from Nuclei},
Nucl. Phys. A \textbf{280}, 330-350 (1977)

\bibitem{Ozawa:2000gx}
A.~Ozawa, O.~Bochkarev, L.~Chulkov, D.~Cortina,
H.~Geissel, M.~Hellstrom, M.~Ivanov, R.~Janik, K.~Kimura
and T.~Kobayashi, \textit{et al.}
{\em Measurements of interaction cross sections
for light neutron-rich nuclei at relativistic energies
and determination of effective matter radii},
Nucl. Phys. A \textbf{691}, 599-617 (2001)

\bibitem{Ozawa:2001hb}
A.~Ozawa, T.~Suzuki and I.~Tanihata,
{\em Nuclear size and related topics},
Nucl. Phys. A \textbf{693}, 32-62 (2001)

\bibitem{Hassan:2015dfa}
M.~A.~M.~Hassan, M.~S.~M.~Nour El-Din, A.~Ellithi,
E.~Ismail and H.~Hosny,
{\em The effect of halo nuclear density on reaction
cross-section for light ion collision},
Int. J. Mod. Phys. E \textbf{24}, 1550062 (2015).

\bibitem{Al-Khalili:1996ugd}
J.~S.~Al-Khalili and J.~A.~Tostevin,
{\em Matter radii of light halo nuclei},
Phys. Rev. Lett. \textbf{76}, 3903-3906 (1996)
[arXiv:nucl-th/9604033 [nucl-th]].

\bibitem{Al-Khalili:1996our}
J.~S.~Al-Khalili, J.~A.~Tostevin and I.~J.~Thompson,
{\em Radii of halo nuclei from cross section measurements},
Phys. Rev. C \textbf{54}, 1843-1852 (1996)

\end{thebibliography}
\end{document}